\documentclass[12pt,fleqn]{article}
\usepackage{graphics}
\usepackage{psfrag}
\usepackage{a4}

\begin{document} \openup6pt

\title{\Large{Higher dimensional inhomogeneous  dust collapse
and cosmic censorship}}
\author{S.G.~Ghosh\thanks{Author to whom all correspondence should be
directed; E-mail:
sgghosh@hotmail.com}   \\
 Department of Mathematics, Science College, Congress Nagar, \\
Nagpur - 440 012, India
\and and \\
A.~Beesham\thanks{E-mail: abeesham@pan.uzulu.ac.za} \\
Department of  Mathematical Sciences, University of Zululand, \\
Private Bag X1001, Kwa-Dlangezwa 3886, South Africa}

\date{}

\maketitle

\begin{abstract}
We investigate the occurrence and nature of a naked singularity in
the gravitational collapse of an inhomogeneous dust cloud
described by  higher dimensional Tolman-Bondi space-times.  The
naked singularities are found to be gravitationally strong in the
sense of Tipler.  Higher dimensions seem to favour black holes
rather than naked singularities.
\end{abstract}
{\bf Key Words:} Naked singularity, cosmic censorship, higher dimensions \\
{\bf PACS number(s):} 04.70.Dw; 04.70.Bw; 04.20.Cv

\section{Introduction}
The work on  inhomogeneous spherical collapse was pioneered in
1934 by Tolman \cite{rt} who found the metric for space-time with
dust which was extended by Bondi \cite{hb}.  Since then the
metric, known as the Tolman-Bondi metric, is extensively used to
study the formation of naked singularities in spherical collapse.
It is seen that the Tolman-Bondi metric admits both naked and
covered singularities depending upon the choice of initial data
and that there is a smooth transition from one phase to the other
\cite{es}-\cite{djd}. However, according to  the cosmic censorship
conjecture (CCC) \cite{rp}, the singularities that appear in
gravitational collapse are always surrounded by an event horizon.
Moreover, according to the strong version of the CCC, such
singularities are not even locally naked, i.e., no non-spacelike
curve can emerge from such singularities (see \cite{r1} for
reviews on the CCC).   The CCC has as yet no precise
 mathematical formulation or proof for either version. Hence the CCC remains
 one of the most important unsolved problems in general relativity and
 gravitation theory today.  Consequently examples which appear to violate the CCC are
 important and they are an important tool to study this important issue.

Motivated by the development of superstring and other field
theories, there is a renewed interest towards models with extra
dimensions from the viewpoint of both cosmology \cite{rs1} and
gravitational collapse \cite{hd}.  In this context, one question
which is important and naturally arises is, would the examples of
naked singularities in four-dimensional (4D) spherical
gravitational collapse go over to higher dimensional (HD)
space-time or not? If the answer to this question is yes, then a
related question is whether the dimensionality of space-time has
any effect on the formation and nature of the singularity.  It is
therefore interesting to study gravitational collapse of
inhomogeneous dust in HD space-times. There is some literature on
HD inhomogeneous dust collapse \cite{bsc} from the viewpoint of
CCC. However, these studies are restricted to 5D space-time and
cannot reduce to the 4D case.

In this paper we generalize previous studies of 4D spherical
inhomogeneous gravitational collapse  to $(n+2)$-dimensional
space-times (where $n\geq2$) and show that gravitational collapse
of higher-dimensional space-times gives rise to shell focusing
naked singularities which are also gravitationally strong. The
conventional picture of  spherical gravitational collapse
described by 4D Tolman-Bondi space-time can be recovered from our
analysis. This would be discussed in  section III, which will be
followed by our concluding remarks. In  section II we review HD
Tolman-Bondi solutions.

The metric considered here is the generalization of the
Tolman-Bondi metrics. We have used units which fix the speed of
light and the gravitational constant via $8\pi G = c^4 = 1$.

\section{Higher Dimensional Tolman-Bondi Solution}

The HD Tolman-Bondi metric is given in co-moving coordinates as:
\begin{equation}
ds^2 = dt^2 -  e^{\lambda(r,t)}dr^2 - R^2(t,r) d \Omega^2
\label{eq:me}
\end{equation}
where
\begin{equation}
d\Omega^2 = \sum_{i=1}^{n+1} \left[ \prod_{j=1}^{i-1}
sin^2(\theta_j) \right] d \theta_i^2 = d \theta_1^2+ sin^2
\theta_1 (d \theta_2^2+sin^2 \theta_2 (d\theta_3^2+\;.\; .\; .\; +
sin^2 \theta_{n-1} d\theta_n^2) \label{eq:ns }
\end{equation}
is the metric on an $n$-sphere and $n=D-2$ (where $D$ is the total
number of dimensions), together with the stress-energy tensor for
dust:
\begin{equation}
 T_{ab} =  \epsilon(t,r) \delta_{a}^t \delta_{b}^t \label{eq:emt}
\end{equation}
where $u_a = \delta_t^a$ is the $(n+2)$-dimensional velocity. The
coordinate $r$ is the co-moving radial coordinate, $t$ is the
proper time of freely falling shells, $R$ is a function of $t$
and $r$ with $R>0$. With the metric (\ref{eq:me}), the independent
non-vanishing Einstein tensor components are
\begin{equation}
G^t_t = \frac{n(n-1)}{2 R^2} (e^{- \lambda} R'^2 - \dot{R}^2 -1) -
\frac{n}{2}\frac{1}{R} ( \dot{R} \dot{\lambda} + e^{-\lambda} R'
\lambda') + n e^{-\lambda} \frac{R''}{R} = \epsilon \label{eq:gtt}
\end{equation}
\begin{equation}
G^r_r = \frac{n(n-1)}{2 R^2} (e^{- \lambda} R'^2 - \dot{R}^2 -1) -
n \frac{\ddot{R}}{R} = 0 \label{eq:grr}
\end{equation}
\begin{eqnarray}
G^{\theta_1}_{ \theta_1} = G^{\theta_2}_{\theta_2} =\; . \; . \;
. \; = G^{\theta_{n}}_{ \theta_{n}} = \frac{(n-1)(n-2)}{2}
\frac{1}{R^2} (e^{- \lambda} R'^2 - \dot{R}^2-1) - \nonumber  \\
\frac{n-1}{2}\frac{1}{R} (\dot{R} \dot{\lambda} + e^{- \lambda}
R' \lambda') - (n-1) \frac{1}{R}(\ddot{R} - e^{- \lambda}
R'')-\frac{1}{2} (\ddot{\lambda}+\frac{\lambda^2}{2}) = 0
\label{eq:gth}
\end{eqnarray}
\begin{equation}
G^t_r = \frac{n}{2R} (2 \dot{R}'-\lambda R') = 0
 \label{eq:gtr}
\end{equation}
where an over-dot and prime denote the partial derivative with
respect to t and r, respectively.

Integration of Eq. (\ref{eq:gtr}) gives
\begin{equation}
e^{\lambda(r,t)} = \frac{R'^2}{1+f(r)} \label{eq:el}
\end{equation}
which can be substituted into Eq. (\ref{eq:grr}) to yield
\begin{equation}
\dot{R}^2 = \frac{F(r)}{R^{(n-1)}} + f(r) \label{eq:fe}
\end{equation}
 The functions  $F(r)$ and  $f(r)$
are arbitrary and are referred as  the mass and energy functions,
respectively. Since in the present discussion we are concerned
with gravitational collapse, we require that $\dot{R}(t,r) < 0$.
The energy density $\epsilon(t,r)$ is calculated as
\begin{equation}
 \epsilon(t,r) = \frac{n F'}{2 R^n R'}  \label{eq:edt}
\end{equation}
 For physical reasons, one assumes that
the energy density $\epsilon$ is everywhere nonnegative. The
special case $f(r)=0$ corresponds to the marginally bound case
which is of interest to us in this paper.  In this case Eq.
(\ref{eq:fe}) can easily be integrated to
\begin{equation}
t - t_{c}(r) = - \frac{2}{(n+1)}\frac{R^{(n+1)/2}}{\sqrt{F}}
\label{eq:ct}
\end{equation}
where $t_{c}(r)$ is an arbitrary function of integration which
represents the proper time for the complete collapse of a shell
with coordinate $r$.  As it is possible to make an arbitrary
re-labelling of spherical dust shells by $r \rightarrow g(r)$,
without loss of generality, we fix the labelling by requiring
that, on the hyper-surface $t = 0$, $r$ coincide with the radius
\begin{equation}
R(0,r) = r              \label{eq:ic}
\end{equation}
This corresponds to the following choice of $t_{c}(r)$
\begin{equation}
t_{c}(r) =  \frac{2}{(n+1)}\frac{r^{(n+1)/2}}{\sqrt{F}}
\label{eq:ct1}
\end{equation}
The central singularity occurs at $r=0$, the corresponding time
being $t=t_0(0)=0$.  We denote by $\rho(r)$ the initial density:
\begin{equation}
\rho(r) \equiv \epsilon(0,r) = \frac{n F'}{2 r^n} \Rightarrow F(r)
= \frac{2}{n} \int \rho(r) r^n dr  \label{eq:Fr}
\end{equation}

It can be seen from Eq. (\ref{eq:edt}) that the density diverges
faster in HD as compared to 4D. Thus given a regular initial
surface, the time for the occurrence of the central
shell-focusing singularity for the collapse developing from that
surface is reduced as compared to the 4D case for the marginally
bound collapse. The reason for this stems from the form of the
mass function in Eq. (\ref{eq:Fr}). In a ball of radius $0$ to
$r$, for any given initial density profile $\rho(r)$, the total
mass contained in the ball is greater than in the corresponding
4D case. In the 4D case, the mass function $F(r)$ involves the
integral $\int \rho(r) r^2 dr$ \cite{r1}, as compared to the
factor
 $r^n$ in the HD case. Hence, there is relatively more mass-energy collapsing
 in the space-time as compared to the 4D case, because of the assumed overall
 positivity of mass-energy (energy condition). This explains why the collapse is
 faster in the HD case.

Clearly, the time coordinate and radial coordinate are,
respectively, in the ranges $ - \infty < t < t_{c}(r)$ and $0
\leq r < \infty$.

\section{Existence and Nature of Naked Singularity}
In the context of Tolman-Bondi space-times, shell crossing
singularities are defined by $R'=0$ and they can be naked.   It
has been shown \cite{rn} that shell crossing singularities  are
gravitationally weak and hence such singularities cannot be
considered
 seriously in the context of the CCC.  On the other hand,
 central shell focusing singularities (characterized by $R=0$ and $R'=0$)
 are also naked and gravitationally strong as well.  Thus, unlike
 shell crossing singularities, shell focusing
 singularities do not admit any metric extension
 through them.  Here we wish to investigate the similar situation
in our HD space-time.
 Christodoulou \cite{dc} pointed out in the 4D case
that the non-central singularities are not naked. Hence, we shall
confine our discussion to the central shell focusing singularity.
Eq. (\ref{eq:ct}), by virtue of Eq. (\ref{eq:ct1}), leads to
\begin{equation}
R^{(n+1)/2} = r^{(n+1)/2} - \frac{(n+1)}{2} \sqrt{F} t
\label{eq:sf}
\end{equation}
and the energy density becomes
\begin{equation}
\epsilon(t,r) = \frac{2n/(n+1)} {\left[ t - \frac{2}{(n+1)}
\frac{r^{(n+1)}}{\sqrt{F}} \right] \left[t - \frac{2 \sqrt{F}}{F'}
r^{(n-1)/2} \right]}
\end{equation}
We are free to specify $F(r)$ and we consider a class of models
which are non self-similar in general, and as a special case, the
self-similar models can be constructed from them. In particular,
we suppose that $F(r) = r^{(n-1)} \zeta(r)$ and $\zeta(0) =
\zeta_{0} > 0 $ (finite). With this choice of $F(r)$, the density
behaves as inversely proportional to the square of time at the
centre, and $F(r) \propto r^{(n-1)}$ in the neighborhood of $r=0$.
For space-time to be self-similar, we require that $\zeta(r) =
const.$ This class of models for 4D space-time is discussed in
refs. \cite{dj}. From Eq. (\ref{eq:edt}) it is seen that the
density at the centre ($r=0$) behaves with time as $\epsilon =
2n/(n+1) t^2$.  This means that the density is finite at any time
$t = t_0 < 0$, but becomes singular at $t=0$.   Thus the
singularity is interpreted as having arisen from the evolution of
dust which had a finite density distribution in the past on an
initial epoch.

We wish to investigate if the singularity, when the central shell
with  co-moving coordinate $(r=0)$ collapses to the centre at
time $t=0$, is naked.  The singularity is naked iff there exists
a null geodesic which emanates from the singularity.  Let
$K^a=dx^a/dk$ be the tangent vector to the radial null geodesic,
where $k$ is an affine parameter.  Then we derive the following
equations
\begin{equation}
\frac{dK^t}{dk} + \dot{R}' K^r K^t = 0 \label{eq:ngf}
\end{equation}
 \begin{equation}
\frac{dt}{dr} =  \frac{K^t}{K^r} = R' \label{eq:ng}
\end{equation}
The last equation, upon using Eq. (\ref{eq:fe}), turns out to be
\begin{equation}
\frac{dt}{dr} = \frac{r^{(n-1)/2} - \frac{ F'}{\sqrt{F}} t}{
\left[ r^{(n+1)/2} - \frac{n+1}{2}
\sqrt{F}\right]^{\frac{n-1}{n+2}} }
 \label{eq:ng1}
\end{equation}
Clearly this differential equation becomes singular at
$(t,r)=(0,0)$. We now wish to put Eq. (\ref{eq:ng1}) in a form
that will be more useful for subsequent calculations. To this
end, we define two new functions $\eta = rF'/F$ and $P = R/r$.
From Eq. (\ref{eq:fe}), for $f=0$, we have $\dot{R}= -
\sqrt{F}/R^{(n-1)}$, and we can express $F$ in terms of $r$ by
$F(r)=r^{(n-1)} \zeta(r)$. Eq. (\ref{eq:ng1}) can thus be
re-written as
\begin{equation}
\frac{dt}{dr} = \left[ \frac{t}{2 r} \eta - \frac{1}{\sqrt{\zeta}}
 \right] \dot{R} \label{eq:ng2}
\end{equation}
It can be seen that the functions $\eta(r)$
and $P(r,t)$ are well defined when the singularity is approached.

Let us  define $X = t/r$ as usual. From the definitions of $P$ and
$X$, and Eq. (\ref{eq:sf}), we can derive the following equation
\begin{equation}
X - \frac{2}{(n+1) \sqrt{\zeta}} = \frac{-P^{(n+1)/2}}{(n+1)
\sqrt{\zeta}} \label{eq:pe2}
\end{equation}
 The nature (a naked singularity or a
black hole) of the singularity can be characterized by the
existence of radial null geodesics emerging from the
singularity.  The singularity is at least locally naked if there
exist such geodesics, and if no such geodesics exist, it is a
black hole. If the singularity is naked, then there exists a real
and positive value of $X_{0}$ as a solution to the algebraic
equation \cite{r1}
\begin{equation}
X_{0} = \lim_{t\rightarrow 0 \; r\rightarrow 0} X =
\lim_{t\rightarrow 0 \; r\rightarrow 0} \frac{t}{r}=
\lim_{t\rightarrow 0 \; r\rightarrow 0} \frac{dt}{dr} = R'      \label{eq:lm1}
\end{equation}

We insert Eq. (\ref{eq:ng2}) into Eq. (\ref{eq:lm1}) and use the
results $\lim_{r\rightarrow 0} \eta = n-1$ and
$\lim_{r\rightarrow 0}\dot{R} = - \sqrt{\lambda}/Q_0^{(n-1)/2}$ to
get
\begin{equation}
X_{0} = \frac{1}{Q_0^{(n-1)/2}} \left[1 - \frac{n-1}{2} X_{0}
\sqrt{\zeta_{0}} \right] \label{eq:pe1}
\end{equation}
where $Q(X) = P(X,0)$. From Eq. (\ref{eq:pe2}) we can find $Q_{0}$
and  then substituting $Q_0$ into Eq. (\ref{eq:pe1}),  we get the
algebraic equation
\begin{equation}
X_0 \left[1 - \frac{(n+1)}{2} \sqrt{\zeta_0} X_0
\right]^{(n-1)/(n+1)} + \frac{(n-1)}{2} \sqrt{\zeta_0} X_0 - 1 =0
\label{eq:pe7}
\end{equation}
which can written as
\begin{equation}
X_0 \left[ 1- \frac{1}{\Theta^0_{(n+2)}} X_0
\right]^{(n-1)/(n+1)} + \frac{(n-1)}{(n+1)\Theta^0_{(n+2)}}  X_0
-1 = 0 \label{eq:pe3}
\end{equation}
where $\Theta_{n+2}$ is defined by $t_c(r)= \Theta_{(n+2)}r$ and
$\Theta^0_{(n+2)} = \Theta_{n+2}(0)$. From Eq. (\ref{eq:pe2}), it
is clear that $0 < X < \Theta_{n+2}$, as $P$ is a positive
function. This algebraic equation governs the behaviour of the
tangent near the singular points.  The central shell focusing is
at-least locally naked (for brevity we address it as naked
throughout this paper) if Eq. (\ref{eq:pe3}) admits one or more
positive roots subject to the constraint that $0 < X_0 <
\Theta_0^{(n+2)}$. The values of the roots give the tangents of
the escaping geodesics near the singularity.  The smallest of
value of $X_0$, say $X_0^s$, corresponds to the earliest ray
escaping from the singularity and is called the Cauchy horizon of
the space-time and  there is no solution in the region $X <
X_0^s$. Thus in the absence of a positive root to Eq.
(\ref{eq:pe3}), the central singularity is not naked because
there are no outgoing future directed null geodesics emanating
from the singularity.

To extract some information from our analysis we start from the
standard 4D Tolman-Bondi case, which has been analyzed earlier by
many authors (see for example, \cite{dj}).  We already know what
happens and we refer the reader to these papers for details.
Setting $n=2$, Eq. (\ref{eq:pe3}) simplifies to
\begin{equation}
X_0 \left[1 - \frac{3}{2} \sqrt{\zeta_0}  X_0 \right]^{1/3} +
\frac{1}{2} \sqrt{\zeta_0} X_0 - 1 = 0 \label{eq:pe4}
\end{equation}
To facilitate comparison with Singh and Joshi \cite{dj}, we
introduce a new variable $y$ defined as
\begin{equation}
y = \sqrt{\zeta_0}  X_0 \nonumber
\end{equation}
After some rearrangement, equation (\ref{eq:pe4}) takes the form
as in Joshi and Singh \cite{dj}:
\begin{equation}
y^3 (y-\frac{2}{3}) - \gamma (y-2)^3 =0 \label{eq:pe5}
\end{equation}
where $\gamma = \zeta^{(3/2)}/12$ and $y < 2/3$. It can be shown
that Eq. (\ref{eq:pe5}) has positive roots, subject to the
constraint that $y<2/3 $ , if $
 \gamma  < \gamma_1 = {26}/{3} - 5 \sqrt{3} \;\approx \; 6.41626 \times
10^{-3}$  or equivalently if $\zeta_0  \leq \zeta_0^C =0.1809$.
For example if $ \zeta_0 = 0.1808$, then the two roots of Eq.
(\ref{eq:pe4}) are $y = 0.530448 \;,\; y =0.541188$ and the
corresponding values of $X_0$ are shown in table-II below.  Eq.
(\ref{eq:pe4}) admits positive real roots for $\gamma < \gamma_2 =
{26}/{3} + 5 \sqrt{3} \;\approx \; 17.32$ also.  The range $
\gamma > \gamma_2$ however, is unphysical because it corresponds
to imaginary $X_0$ \cite{dj}. Thus referring to our above
discussion, collapse leads to a naked singularity for $\gamma <
\gamma_1$ ($\zeta_0 \leq 0.1809$) and to a black hole otherwise.
This agrees with the earlier results \cite{dj}.\\

We now wish to investigate the changes introduced, at least
qualitatively, in the above picture by the introduction of extra
dimensions. To conserve space, we avoid repetition  of the
detailed analysis (being similar to 4D case) and rather summarize
only the main results in the following tables:

\pagebreak \noindent {\bf Table I: The Variation of the
$\zeta_0^C$, $\Theta_{n+2}^0 $ and  $X_0$ \\ with $D=n+2$}

\begin{tabular}{|c|c|c|c|c|} \hline
  $n$ & $D$ & $\zeta_0^C$& $\Theta_{n+2}^0$  & Equal tangents ($X_0$) \\  \hline
  $2$ & $4$ & $0.18091652297$   &$1.56736$  &$1.25992$ \\  \hline
  $3$ & $5$ & $0.090169943745$  &$1.6651$   &$1.27202$ \\  \hline
  $4$ & $6$ & $0.056372858735$  &$1.68471$  &$1.26751 $ \\  \hline
  $5$ & $7$ & $0.039372532807$  &$1.67989$  &$1.25992 $ \\  \hline
  $6$ & $8$ & $0.0182853577435$ &$2.11291$  &$1.45161 $ \\  \hline
\end{tabular}
\\

\noindent
 {\bf Table II: The two tangents for a  $\zeta
< \zeta_0^C$, \\ in the different space-times.}

\begin{tabular}{|c|c|c|c|} \hline
  $n$ & $D$ & $ \zeta_0 < \zeta_0^C$ & Two tangents ($X_0$)  \\ \hline
  2 & 4 & 0.1808 & $1.2476,  \; 1.27277$\\   \hline
  3 & 5 & 0.0900 & $1.25,    \; 1.29571$ \\  \hline
  4 & 6 & 0.0563 & $1.24925, \; 1.28692$ \\  \hline
  5 & 7 & 0.0393 & $1.23863, \; 1.28284$ \\  \hline
  6 & 8 & 0.1820 & $1.40558, \; 1.50272$ \\  \hline
\end{tabular}

The quantities $\zeta_0$, $\Theta_{n+2}$ and $X_0$ depend on the
dimension of the space-time.  Thus it follows that the singularity
is naked if $\zeta_0 \leq \zeta_0^C$ (of course the roots are
subject to the constraint that $X_0 < \Theta_{n+2}$). On the other
hand, if the inequality is reversed, no naked singularity forms
and gravitational collapse of the dust collapse will result in a
black hole. The quantity $\zeta^C_0 $ can be called the critical
parameter as at $\zeta^C_0= \zeta_0 $, a transition occurs and the
end state of collapse turns from a naked singularity to a black
hole.  It is interesting to note that $\zeta^C_0$ decreases as one
introduces extra dimensions, whereas $\Theta_{n+2}$ increases. We
have plotted a graph of critical parameter $\zeta^C_0$ against
dimension D (see Figure I). Thus we can say that the formation of
a black hole is facilitated with introduction of the extra
dimensions or, in other words, the naked singularity spectrum gets
continuously covered in higher dimensional space-times. Thus it
appears that the singularity will be completely covered for very
large dimensions of the space-time. Our results are in agreement
with earlier work \cite{bsc} on 5D spherical inhomogeneous dust
collapse. A similar situation also occurs in HD radiation collapse
\cite{ns}.

\input{epsf}
\begin{figure}
\begin{center}
\psfrag*{x_c}{$\zeta_0^C$}
\includegraphics{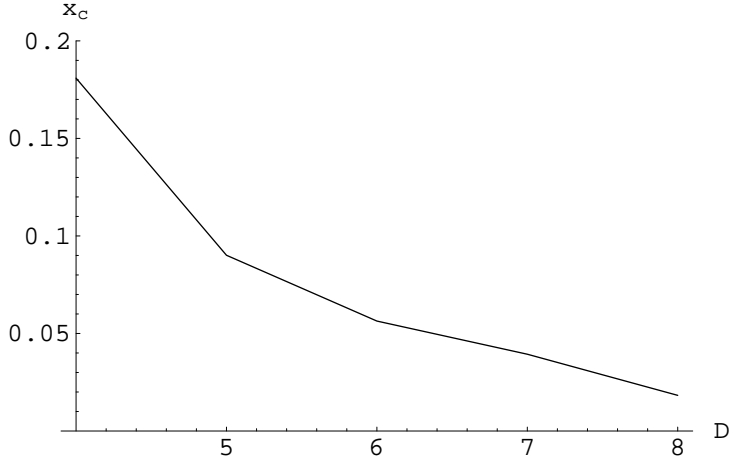}
\caption{Plot of $\zeta_0^C$ versus  $D$}
\end{center}
\end{figure}

\subsection{Strength of Naked Singularity}
Finally, we need to determine the curvature strength of the naked
singularity which is an important aspect of a singularity
\cite{ft1}. There has been an attempt to relate the strength of a
singularity to its stability \cite{djd}.  A singularity is
gravitationally strong or simply strong if it destroys by crushing
or stretching any object which fall into it and weak if no object
which falls into the singularity is destroyed in this way. It is
widely believed
 that a space-time does not admit an extension through a singularity if it is
a strong curvature singularity in the sense of Tipler \cite{ft}.
Clarke and Kr\'{o}lak \cite{ck} have shown that a sufficient
condition for a strong curvature singularity as defined by Tipler
\cite{ft} is that for at least one non-space like geodesic with
affine parameter $k$, in the limiting approach to the
singularity, we must have
\begin{equation}
\lim_{k\rightarrow 0}k^2 \psi =
\lim_{k\rightarrow 0}k^2 R_{ab} K^{a}K^{b} > 0 \label{eq:sc}
\end{equation}
where $R_{ab}$ is the Ricci tensor.  Our purpose here is to
investigate the above condition along future directed radial null
geodesics which emanate from the naked singularity. Now
(\ref{eq:sc}), with the help of Eq. (\ref{eq:emt}) and using $P =
R/r$, can be expressed as
\begin{equation}
\lim_{k\rightarrow 0}k^2 \psi = \lim_{k\rightarrow 0}
\frac{nF'}{2r^{(n-2)}P^{(n-2)} R'} \left[ \frac{k K^t}{R}
\right]^2 \label{eq:sc1}
\end{equation}
In order to evaluate the above limit, we note that the tangent
$K^t$ blows up in the limiting approach to the naked singularity.
Using L'Hospital's rule, we find that
\begin{equation}
\lim_{k\rightarrow 0} k^2 \psi= \lim_{k\rightarrow 0} \frac{k}{R
\frac{1}{K^t}}=\lim_{k\rightarrow 0}  \frac{1}{\frac{R
\dot{R}'}{R'} + \dot{R} +1} \label{eq:sc2}
\end{equation}
The quantity $\dot{R}'$, from  Eq. (\ref{eq:fe}), can be
calculated as:
\begin{equation}
\dot{R}'= - \frac{1}{R} \left[ \frac{\eta \sqrt{\lambda}}{2
P^{\frac{(n-1)}{2}-1}} + \frac{n-1}{2}\dot{R} R' \right]
\label{eq:sc3}
\end{equation}
Now Eq. (\ref{eq:sc}), because of Eqs. (\ref{eq:sc1}),
(\ref{eq:sc2}) and (\ref{eq:sc3}), and using the fact that
$F/r^{(n-1)}$ and $R'$ tend to finite values $\lambda_0 (n-1)$ and
$X_0$ respectively, yields
\begin{equation}
\lim_{k\rightarrow 0}k^2 \psi = \frac{n \lambda_0 (n-1)}
{\left[[(n-3)X_0 - (n-1) Q_0] \sqrt{\lambda_0}+2 X_0
Q_0^{(n-1)/2} \right]^2} > 0 \label{eq:sc4}
\end{equation}
Thus along radial null geodesics coming out from singularity
$\lim_{k\rightarrow 0}k^2 \psi$ is finite and hence the strong
 curvature condition is satisfied.
\section{Concluding remarks}
The Tolman-Bondi metric in the 4D case is extensively used for
studying the formation of naked singularities in spherical
gravitational collapse. Indeed, both analytical
\cite{dj}-\cite{jd1} and numerical results \cite{es} in dust
indicate the critical behaviour governing the formation of black
holes or naked singularities.  For further progress towards an
understanding of spherical dust collapse, from the viewpoint of
the CCC, one would like to know the effect of extra dimensions on
the existence of a naked singularity.  The relevant questions
would be, for instance, do such solutions remain naked with the
introduction of extra dimensions? Do they always become covered?
Does the nature of the singularity change?  Our analysis shows
that none of the above hold. Indeed, the gravitational collapse of
inhomogeneous dust in higher dimensional space-time leads to a
strong curvature naked singularity.   Thus extra dimensions cannot
completely cover the naked singularity nor can it affect the
nature of the singularity. However, we showed that the effect of
extra dimensions appears to be a shrinking of the naked
singularity initial data space (of 4D) or an enlargement of the
black hole initial data space. Thus one can say the that the naked
singularity spectrum in 4D case gets covered with the introduction
of extra dimensions in the space-time or one can say that extra
dimensions of the space-time facilitate the formation of black
holes in comparison to naked singularities.

 This generalizes the
previous studies of spherical gravitational collapse in 4D to HD
space-times and when $n=2$, one recovers the conventional 4D
Tolman-Bondi models. Also for $n=3$, our result reduces to those
obtained previously for the 5D case \cite{bsc}. The formation of
these naked singularities violates the strong CCC. We do not claim
any particular physical significance to the HD metric considered.
Nevertheless we think that the results obtained here have some
interest in the sense that they do offer the opportunity to
explore  properties associated with naked singularities which may
be crucial in our understanding of this important problem.
Finally, the result obtained would also be relevant in the context
of superstring theory which is often said to be next "theory of
everything", and for an interpretation of how critical behaviour
depends on the dimensionality of the space-time.

\noindent {\bf Acknowledgment:} SGG  would like to thank the
University of Zululand for hospitality, the NRF (South Africa) for
financial support, Science College, Congress Nagar, Nagpur
(India) for granting leave and UGC, Pune for MRP F. No
23-118/2000 (WRO). The authors would like to thank the
Harish-Chandra Research Institute, Allahabad, India for kind
hospitality while part of this work was being done.

\noindent
\end{document}